\documentclass[epj]{webofc}
\usepackage[varg]{txfonts}    
\usepackage{graphicx}
\usepackage{amsmath}              
\usepackage{amssymb}
\usepackage{amsfonts}
\usepackage{multirow}
\wocname{EPJ Web of Conferences}
\woctitle{ICNFP 2016}
\begin{document}
\selectlanguage{english}
\title{Pair Production of Beyond the Standard Model Higgs Bosons}
%
%

\author{Ramona Gr\"ober \inst{1,2}\fnsep\thanks{\email{ramona.groeber@durham.ac.uk}} 
}

\institute{Institute of Particle Physics Phenomenology, Department of Physics, Durham University, DH1 3LE, UK
\and
INFN, Sezione di Roma Tre, Via della Vasca Navale 84, 00146 Roma, Italy
}

\abstract{%
Higgs pair production is not only sensitive to the trilinear Higgs self-coupling, but it can give access to other anomolous couplings, as e.g.~a novel $hht\bar{t}$ coupling. In Composite Higgs Models, this coupling usually leads to a large increase of the cross section.
In such a framework an interesting question is, whether it might be possible to observe new physics for the first time in Higgs pair production. This question will be addressed by taking into account projected sensitivities for Higgs coupling measurements and for direct searches of new vector-like quarks.
\par
Higher order corrections to Higgs pair production via gluon fusion are sizeable. It is hence not only important to compute them in the Standard Model (SM) but also in its extensions. Here, the computation of the QCD corrections in the SM with dimension 6 operators as well as the SUSY-QCD corrections to Higgs pair production via gluon fusion in the MSSM are presented. 
}
\maketitle
\section{Introduction}
\label{intro}
An experimental verification of the SM electroweak symmetry breaking mechanism requires the determination of the Higgs self-couplings which can be measured in multi-Higgs production processes.\footnote{An indirect way to determine the trilinear Higgs self-coupling is via higher order corrections to single Higgs production, see refs.~\cite{Bizon:2016wgr}. } 
While the the quartic Higgs self-coupling is out of reach of the LHC due to the small signal cross section of the triple Higgs production process \cite{Plehn:2005nk}, the trilinear Higgs self-coupling is accessible in Higgs pair production processes. The Higgs pair production cross section is typically three orders of magnitude smaller than the one of single Higgs production. This makes a measurement very challenging and requires high luminosities. Phenomenological studies point out that the $b\bar{b}\gamma \gamma$ channel is the most promising final state \cite{Baur:2003gp, hhreview}.
First experimental efforts have started, see e.g.~refs.~\cite{CMS:2016vpz}, but are not yet sensitive to the SM cross section. Nevertheless, they can already restrict certain beyond the SM (BSM) scenarios with largely increased cross section. 
\par
The dominant Higgs pair production process is gluon fusion. The process is mediated by triangle and box diagrams of heavy fermions.
Apart from the trilinear Higgs self-coupling that can be measured in this process, new physics can manifest itself in several different ways as e.g.~in terms of coupling modifications to the top quark Yukawa coupling, a new $hht\bar{t}$ coupling \cite{Dib:2005re, hhttcoup}, new particles
in the loop \cite{Dawson:2015oha} or by the exchange of a new resonance \cite{Chen:2014ask}. In particular in models with an exchange of a new resonance or a $hht\bar{t}$ coupling the cross section can be strongly increased. Focusing on the latter case, an interesting question to ask is whether the modification can be so large that new physics might be observed for the first time in Higgs pair production. In this contribution, I will address this question in terms of Composite Higgs Models by taking into account projected sensitivities on Higgs couplings and direct searches for vector-like quarks.
\par
As for single Higgs production, the higher order corrections to Higgs pair production are rather large and nearly double the cross section. It is hence desirable to include higher order corrections also for beyond the SM (BSM) studies. Here, I will discuss the impact of the higher order corrections in the SM extended with dimension (dim) 6 operators \cite{Grober:2015cwa} and in the minimal supersymmetric extension of the SM (MSSM) \cite{Agostini:2016vze}. 
\par
After reviewing the theoretical status of the SM calculation of the gluon fusion process in section~\ref{SM}, pair production of BSM Higgs bosons will be discussed with a special focus on the question whether new physics might be for the first time observed in Higgs pair production (section~\ref{BSM}) and on higher order corrections to BSM Higgs pair production (section~\ref{NLOBSM}).
\section{Higgs pair production via gluon fusion in the Standard Model}
\label{SM}
The leading order (LO) SM cross section in full top quark mass dependence is known since the late 80's \cite{Glover:1987nx}. Since the process is already at the LO mediated by one-loop triangle and box diagrams, the computation of the next-to-leading order (NLO) corrections involves two-loop diagrams with multiple scales. This makes a full computation of the NLO corrections technically very difficult. 
A useful approximation to compute higher order corrections is the infinite top mass limit. In ref.~\cite{Dawson:1998py} this limit has been employed to compute the NLO cross section. Whereas in single Higgs production the infinite top mass limit works rather well, for Higgs pair production it is only valid if the invariant mass of the Higgs boson pair $m_{hh}$ is much smaller than twice the top quark mass $m_t$, $m_{hh}\ll 2\, m_t$.
The results of ref.~\cite{Dawson:1998py} have been improved by factoring out the exact Born cross section. In ref.~\cite{Grigo:2013rya, Grigo:2015dia} higher terms in the expansion in the large top mass have been computed; in ref.~\cite{Degrassi:2016vss} analytic results for this expansion have been presented. The full real radiation corrections have been computed in ref.~\cite{Frederix:2014hta}. 
Recently, the NLO results in full top mass dependence were given in refs.~\cite{Borowka:2016ehy}. It turned out that compared to the Born improved NLO cross section in the $m_t\to \infty$ limit, the NLO cross section in the full top mass dependence is reduced by approximately $14\%$.
\par
The next-to-next-to leading order (NNLO) corrections in the infinite top mass limit have been presented in ref.~\cite{Grigo:2014jma}, and in ref.~\cite{Grigo:2015dia} higher order terms in an expansion in small external momenta have been computed. The differential cross section at NNLO has been given in \cite{deFlorian:2016uhr}.
Threshold resummation at next-to-next-to-leading logarithmic accuracy computed in the heavy quark mass limit further increases the cross section \cite{deFlorian:2015moa}. Resummation of the Higgs transverse momentum spectrum at next-to-leading logarithmic accuracy in full top quark mass dependence has been performed in ref.~\cite{Ferrera:2016prr}.
The QCD corrections are typically doubling the cross section. The size of the cross section and the associated uncertainties are listed in detail in ref.~\cite{deFlorian:2016spz}.
\section{Can new physics be seen for the first time in Higgs pair production?}\label{BSM}
In this section I will address the question whether the new physics can be seen for the first time in Higgs pair production, meaning whether it could be that we miss new physics in direct searches or indirectly, by e.g. Higgs coupling measurements, and nevertheless see a sizeable deviation in Higgs pair production. This discussion is based on ref.~\cite{Grober:2016wmf}.
\par
The question has to be addressed in concrete models. In case of a new resonance that decays predominately into Higgs bosons a large increase of the cross section can be expected and the likely answer to this question is yes. I will instead consider another case without a new resonance, but with a new $hht\bar{t}$ coupling. Types of models where such a coupling emerges are Composite Higgs Models \cite{hhttcoup}.
\par
The idea of Composite Higgs Models is, that the Higgs boson arises as a pseudo-Goldstone boson of a breaking of a global symmetry at a scale $f$. Effectively the model can be described by a non-linear $\sigma$ model. The minimal (custodial symmetric) model is based on a global SO(5) symmetry that is broken to a SO(4). The two-derivative Lagrangian then reads
\begin{equation}
\mathcal{L}=\frac{f^2}{2} (D_{\mu}\Sigma) (D^{\mu}\Sigma)^T \hspace*{0.5cm}\text{with} \hspace*{0.5cm} \Sigma=exp(-i \sqrt{2} T^{\hat{a}} h^{\hat{a}}(x))\, ,
\end{equation} 
where $T^{\hat{a}}$ with $\hat{a}=1 ... 4$ are the generators of the SO(5)/SO(4) coset and $h^{\hat{a}}(x)$ are the four Goldstone fields (the three Goldstone bosons associated with the longitudinal degrees of freedom of the SM gauge bosons and the Higgs boson). From the covariant derivatives $D_{\mu}\Sigma$ the coupling of the Higgs boson to the gauge bosons can be obtained. They are modified compared to the SM by a parameter $\xi=v^2/f^2=\sin \langle H \rangle/f$, with  $v$ denoting the vacuum expectation value of the spontanous breaking of the electroweak SU(2)$\times$U(1) symmetry. The Higgs self-couplings and the Higgs fermion couplings depend on the fermion embedding into the symmetry group.  
In the following, I will consider three models that differ by their fermion embedding, the MCHM4 \cite{Agashe:2004rs}, the MCHM5 and the MCHM10 \cite{Contino:2006qr}. For the first two models only pure Higgs non-linearities will be considered. A table showing the modifications of the couplings with respect to the SM can be found in ref.~\cite{Grober:2016wmf}. All couplings are completely determined once a value for $\xi$ is chosen. Both models have a $hht\bar{t}$ coupling.
\par 
Whenever referred to the MCHM10, an explicit set of new vector-like fermions is introduced in a $\bf{10_{2/3}}$ of SO(5). Details on the model can be found in ref.~\cite{Gillioz:2013pba}. The model has three additional parameters (mass of the $\bf{10_{2/3}}$, coupling of the $\bf{10_{2/3}}$ with the Goldstone field and a mixing angle) with respect to the MCHM4 and the MCHM5. The $\bf{10_{2/3}}$ contains also bottom partners. This will lead to a dependence of the $hb\bar{b}$ coupling and the $gg\to h$ and $h\to \gamma \gamma$ rates on all model parameters.\footnote{This has to be seen in contrast to the MCHM5 with a $\bf{5_{2/3}}$ of vector-like fermions where $gg\to h$ and $h\to \gamma\gamma$ rates basically only depend on $\xi$ \cite{Falkowski:2007hz}.} 
\par
In order to answer the initial question, we will have to assume that new physics will not be discovered at any point before the LHC will be sensitive to the SM Higgs pair production cross section. This means for the models that are simply modified by Higgs non-linearities that Higgs coupling modifications need to be below the experimental sensitivity for an integrated luminostiy of $L= 300 \text{ fb}^{-1}$ (or  $L= 3000 \text{ fb}^{-1}$ respectively). Projected sensitivities for the coupling measurements can be found in ref.~\cite{Englert:2014uua}. Two final states of the Higgs boson pair will be considered, $b\bar{b}\tau^+\tau^-$ and $b \bar{b} \gamma \gamma$ with the acceptances $A$ estimated from ref.~\cite{hhreview}. Then, the model is said to be distinguishable from the SM if
\begin{equation}
 S_{SM}+3 \sqrt{S_{SM}}\leq S\hspace*{1cm}\text{or}\hspace*{1cm}  
S_{SM}-3\sqrt{S_{SM}}\geq S \label{crit}
\end{equation}
where \begin{equation}
S=\sigma\cdot BR \cdot A\cdot L \;,
\end{equation}
and $BR$ is the respective branching ratio to $b\bar{b}\gamma\gamma$ or $b\bar{b}\tau^+\tau^-$ obtained with a modified version of {\tt HDECAY} \cite{Djouadi:1997yw}. 
\begin{table}
\begin{center}
  \begin{tabular}{|cc|cccc|}\hline
  \multicolumn{2}{|c|}{}  & $\sigma_{b\bar{b}\gamma\gamma}$ [fb] & 
$\Delta_{3\sigma}$ &$\sigma_{b\bar{b}\tau^+ \tau^-}$ [fb]& $\Delta_{3\sigma}$ \\ \hline
\multirow{1}{*}{MCHM4}&$\xi=0.076$ (LHC300)&$0.114$&no&
 3.13&no\\
&$\xi=0.051$ (LHC3000)&0.112&no&
3.07&no\\ \hline
\multirow{1}{*}{MCHM5}&$\xi=0.068$ (LHC300)&$0.175$&no&
 3.96&no\\
&$\xi=0.015$ (LHC3000)&$0.119$&no&
 3.14&no\\ \hline
\end{tabular}
\end{center}
\caption{Values of the cross section times branching ratio in the MCHM4 and 
MCHM5 for the projected values on the sensitivity of Higgs coupling measurements at $L=300\;\text{fb}^{-1}$ and 
$L=3000\;\text{fb}^{-1}$ of ref.~\cite{Englert:2014uua}. In the 4th or 6th row, 
respectively, a "no" indicates that the deviation from the SM rate is below $3\sigma$.}\label{tab:MCHM45}
\end{table}  
In table \ref{tab:MCHM45} the NLO cross section is given for the two final states $b\bar{b}\gamma\gamma$ and $b\bar{b}\tau^+\tau^-$ for the maximal possible values of $\xi$ that lead to Higgs coupling modifications below the expected sensitivity at $L=300\text{ fb}^{-1}$ or $L=3000\text{ fb}^{-1}$, respectively. As can be inferred from the table, in all cases the cross section does not deviate by more than 3$\sigma$ from the SM signal rate.

\par
This indicates that for the considered simple models where the deviations are driven by just one parameter we cannot hope to see any deviations from the SM for the first time in Higgs pair production. 
This analysis can be performed now also for the MCHM10 with fermionic top and bottom partners where the larger parameter space allows for more freedom. In addition to the projected sensitivities on the Higgs couplings, one also needs to consider now that the vector-like fermions can be discovered in direct searches. The lightest new fermion in this model is typically a 5/3 charged fermion, denoted by $\chi$. The reach of the indirect searches can be estimated from the excluded cross section limit at 8 TeV \cite{Aad:2015mba} with
\begin{equation}
r = \sqrt{\frac{\sigma_{BKG}(14
\text{ TeV})}{\sigma_{BKG}(8\text{ TeV})}
\frac{L_{LHC8}}{L_{LHC14}}
} \;,
\end{equation}
where $\sigma_{BKG}$ denotes the cross section of the dominant background $t\bar{t}W^{\pm}$.
The reach on the lightest 5/3 charged fermion can then be estimated to $m_{\chi}=1.37\text{ TeV}$ at $L=300\;\text{fb}^{-1}$ and $m_{\chi}=1.55\text{ TeV}$ at $L=3000\;\text{fb}^{-1}$.
\par
In a scan over the parameter space only points that
are compatible with the estimated sensitivities on the Higgs coupling measurements and the indirect searches are kept. In addition we only generate points that are compatible with electroweak precision tests (EWPTs) at 99\% C.L. \cite{Gillioz:2013pba} and $|V_{tb}|>0.92$ \cite{Chatrchyan:2012ep}.
The Higgs pair production cross section was computed using the NLO QCD corrections as given in \cite{Grober:2016wmf}.
\begin{figure}[t]
\begin{center}
 \hspace*{-0.8cm}\includegraphics[width=7.5cm]{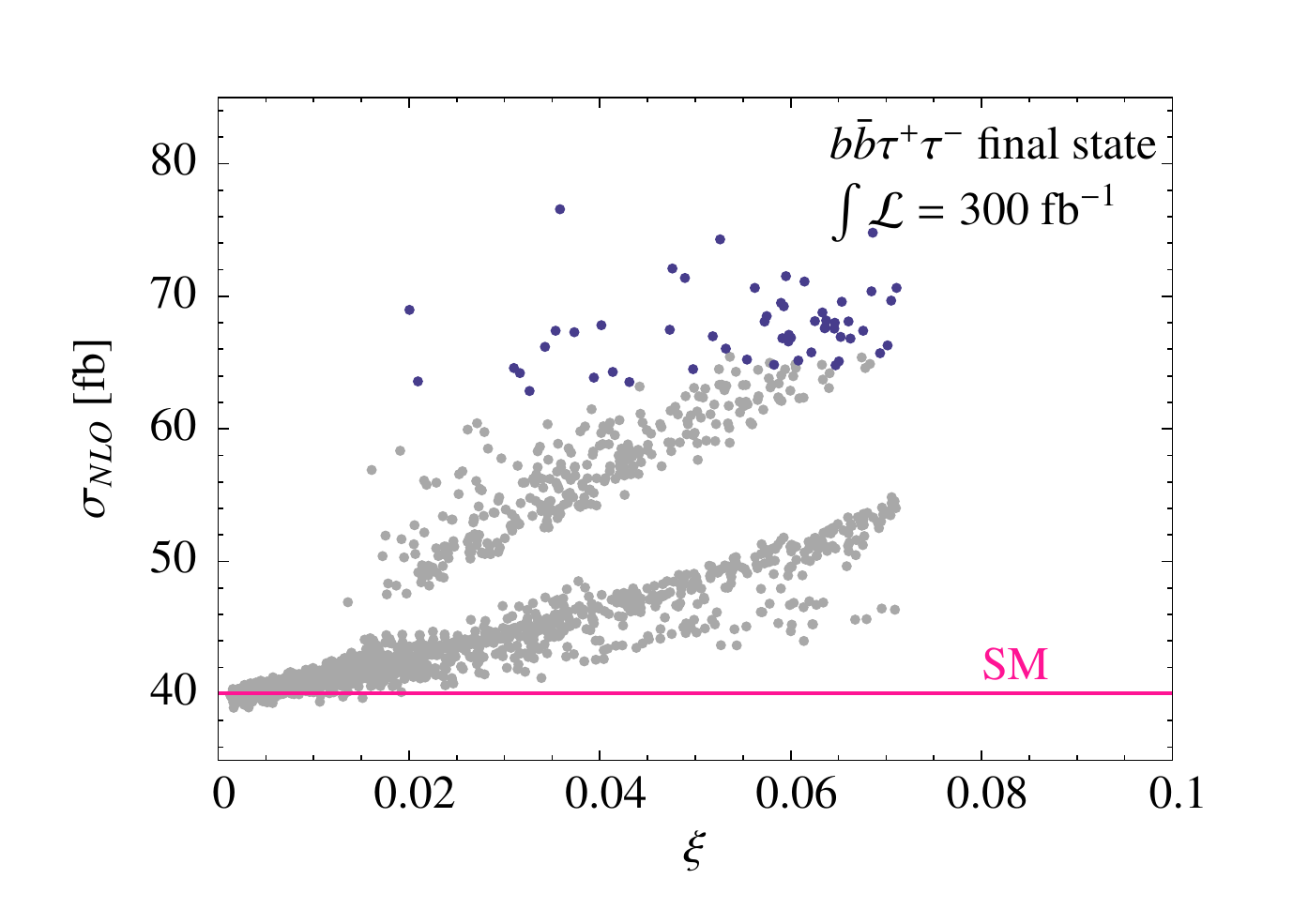}
 \hspace*{-0.2cm}
 \includegraphics[width=7.5cm]{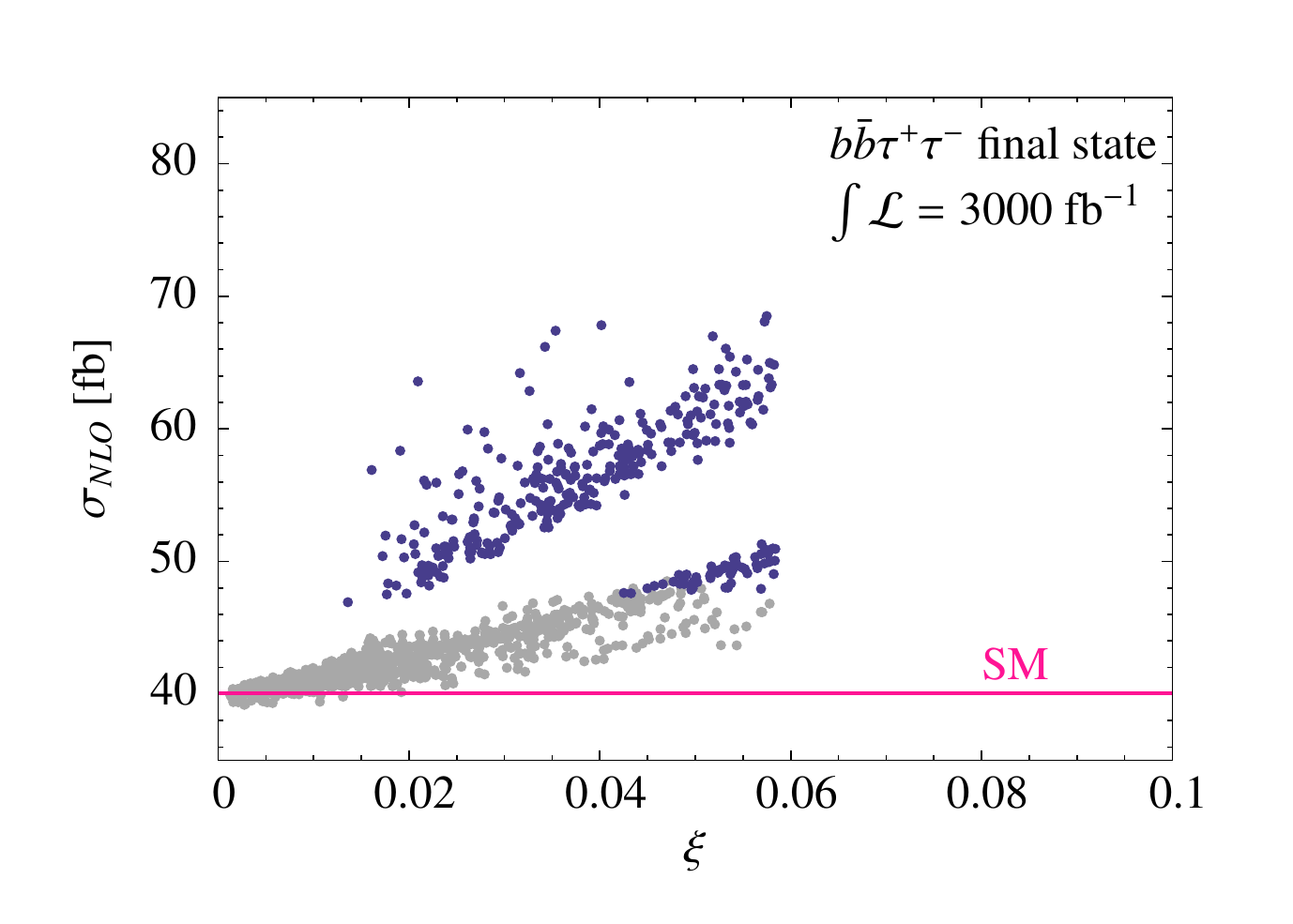}\hspace*{-0.6cm}\\ 
\vspace*{-0.5cm}
  \hspace*{-1.3cm} \includegraphics[width=7.5cm]{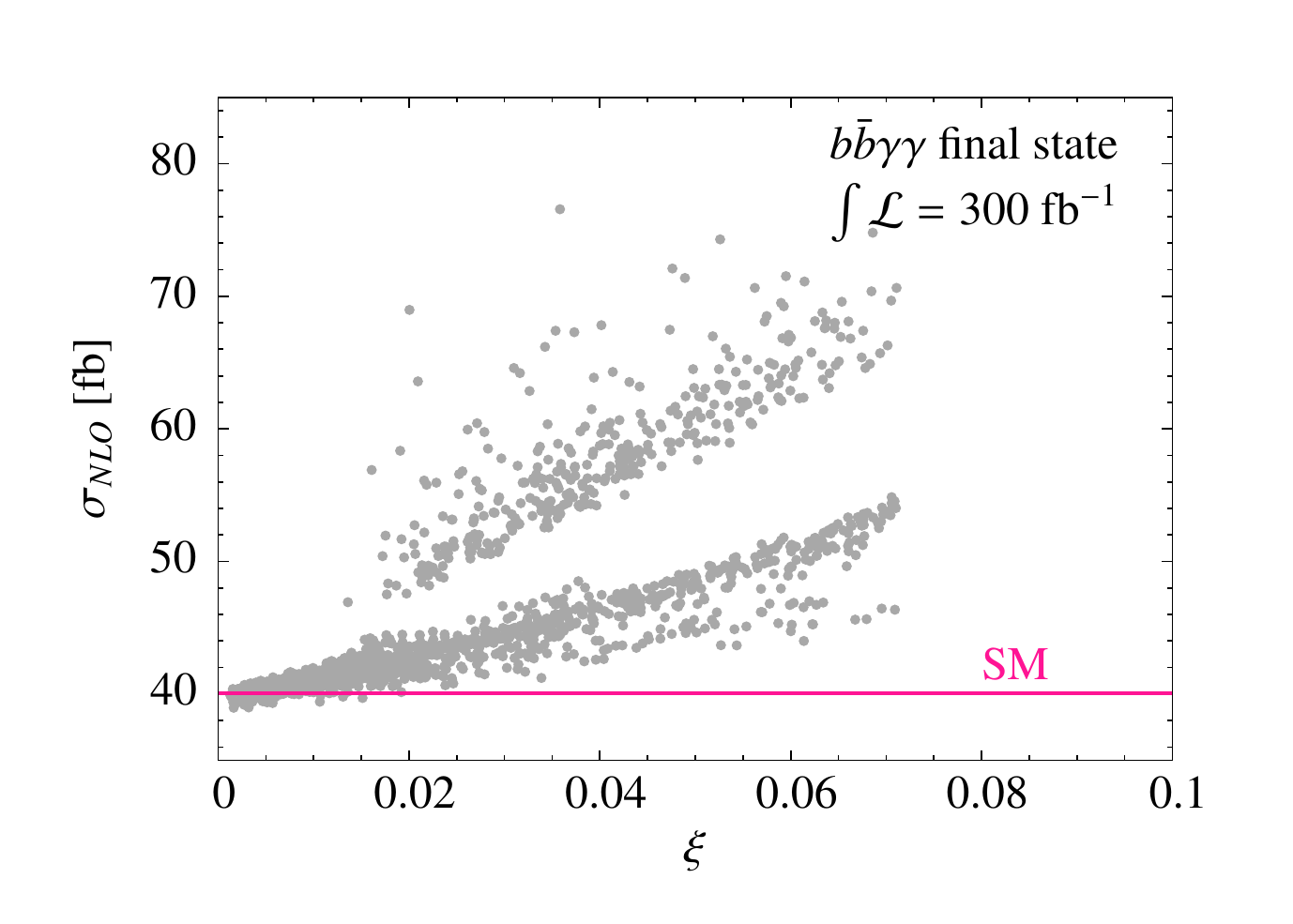}
 \hspace*{-0.2cm}
 \includegraphics[width=7.5cm]{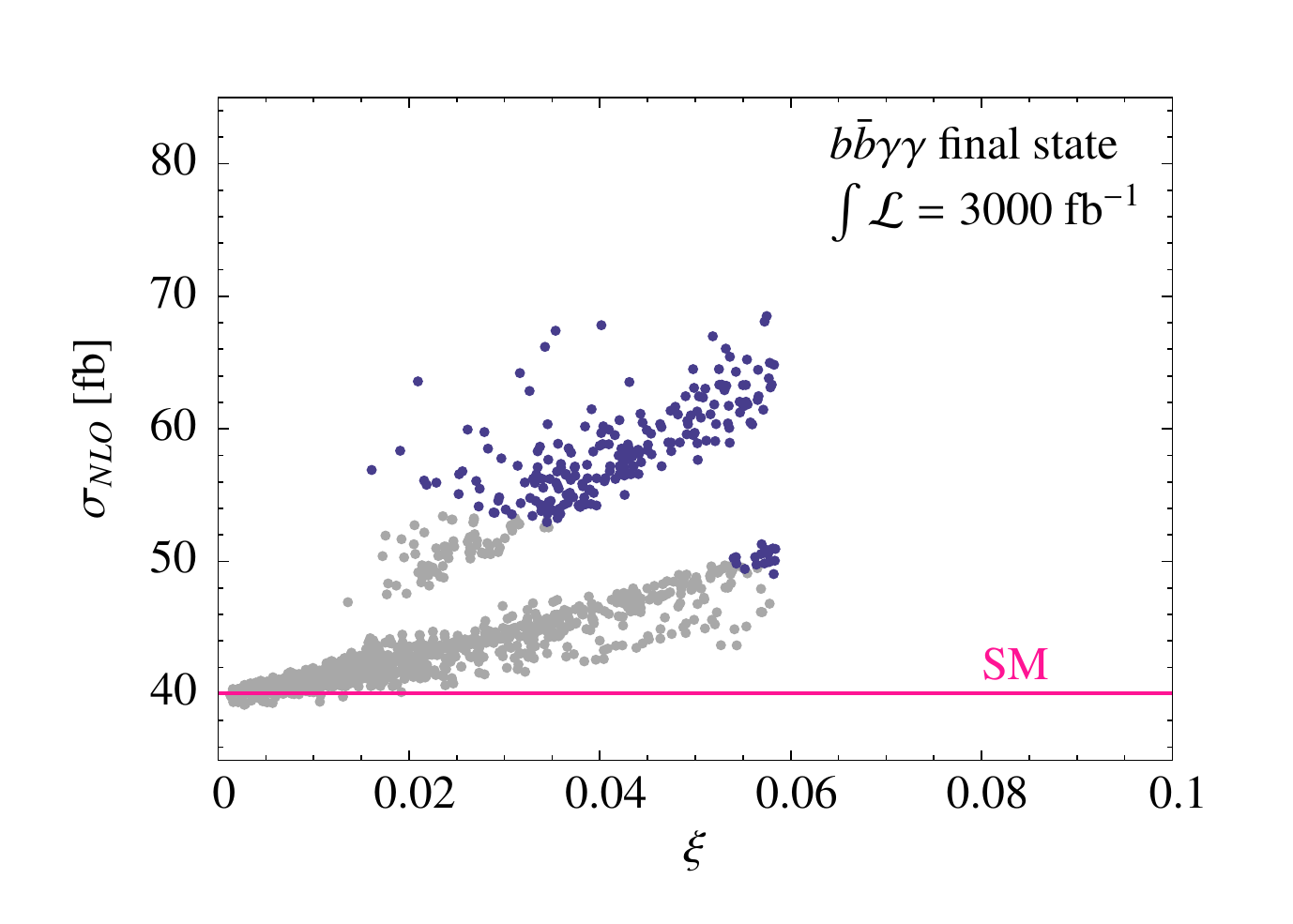}\hspace*{-0.5cm}
 \end{center}
 \vspace*{-0.5cm}
 \caption{The NLO gluon fusion cross section to a Higgs boson pair in
   the MCHM10 for a scan over the parameter space. All
   points of the scan pass the EWPTs, fulfill the projected direct search limits for new fermionic
   resonances at the LHC and allow only for deviations in the Higgs boson
   couplings that are smaller than the expected sensitivity at the
   LHC for the respective luminosity. The blue 
   points indicate that the MCHM10 cross
   section can be distinguished from the SM one at 3$\sigma$ in the 
$b\bar{b}\tau^+ \tau^-$ final state (upper) and the $b\bar{b}\gamma\gamma$ 
final state (lower) at an integrated luminosity $\int \mathcal{L}=300\text{ fb}^{-1}$
(left)  and $\int \mathcal{L}=3000\text{ fb}^{-1}$ (right), whereas the grey points cannot 
be distinguished from the SM at 3$\sigma$. The pink line is the SM
prediction for the gluon fusion cross section at NLO. \label{fig-1}}
\end{figure}
The results can be found in fig.~\ref{fig-1}. As can be inferred from the plots, for $L=300\text{ fb}^{-1}$
only in the $b\bar{b}\tau^+\tau^-$ certain parameter points can be distinguished from the SM at $3 \sigma$, whereas for $L=3000\text{ fb}^{-1}$
in both of the final states there are parameter points that can be distinguished at $3 \sigma$ from the SM cross section.

\section{Higher order corrections to beyond the Standard Model Higgs pair production}\label{NLOBSM}
In this section NLO QCD corrections to BSM Higgs pair production will be discussed: for the SM with effective dim 6 operators~\cite{Grober:2015cwa} in subsection~\ref{dim6} and the SUSY-QCD corrections in the MSSM~\cite{Agostini:2016vze} in subsection~\ref{MSSM}. Reference \cite{Agostini:2016vze} also gives results for the NMSSM, which will however not be discussed here. Furthermore, NLO QCD corrections to Higgs pair production via gluon fusion to the singlet extension of the SM are given in ref.~\cite{Dawson:2015haa}, for the two-Higgs doublet model in ref.~\cite{Hespel:2014sla} and for Composite Higgs Models with and without fermion partners in ref.~\cite{Grober:2016wmf}. 
\subsection{The SM with dim 6 operators}\label{dim6}
The non-linear effective Lagrangian with the relevant dim 6 operators that affect the Higgs pair production cross section are
\begin{equation}
\Delta {\cal L}_{\text{non-lin}} \supset - m_t \bar{t}t \left(c_t
  \frac{h}{v} + c_{tt} \frac{h^2}{2 v^2} \right) 
-c_3 \, \frac{1}{6} \left( \frac{3 M_h^2}{v} \right) h^3 + \frac{\alpha_s}{\pi} G^{a\, \mu\nu}
G_{\mu\nu}^a \left( c_g \frac{h}{v} + c_{gg}\frac{h^2}{2 v^2}
\right) \, , \label{effop}
\end{equation}
with $G^a_{\mu\nu}$ denoting the gluon field strength tensor, $M_h$ the Higgs boson mass and $\alpha_s$ the strong coupling constant.
\par
In ref.~\cite{Grober:2015cwa} the NLO QCD corrections for Higgs pair production via gluon fusion allowing for all of the operators of eq.~\eqref{effop} were computed in the infinite top mass limit following ref.~\cite{Dawson:1998py}. The effective operators enter in the Born cross section and in the reducible two-loop contributions that are given by double triangle integrals. In addition, it has to be taken into account that the operators with Wilson coefficients $c_g$ and $c_{gg}$ do not contribute to the effective gluon coupling at $\mathcal{O}(\alpha_s^2)$.
\par
\begin{figure}\hspace*{-0.9cm}
\includegraphics[width=8.3cm]{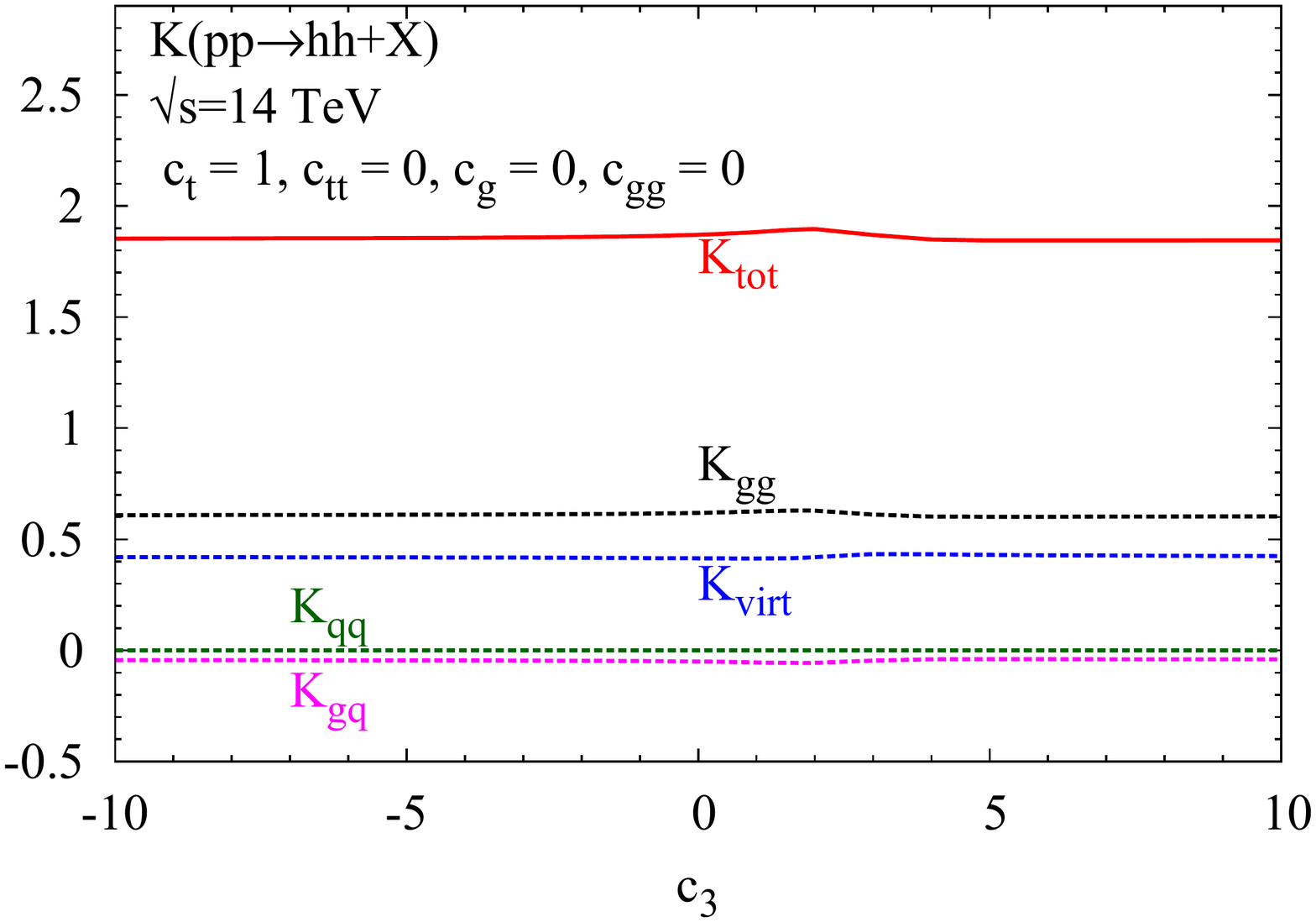}\hspace*{-0.9cm}
\includegraphics[width=8.3cm]{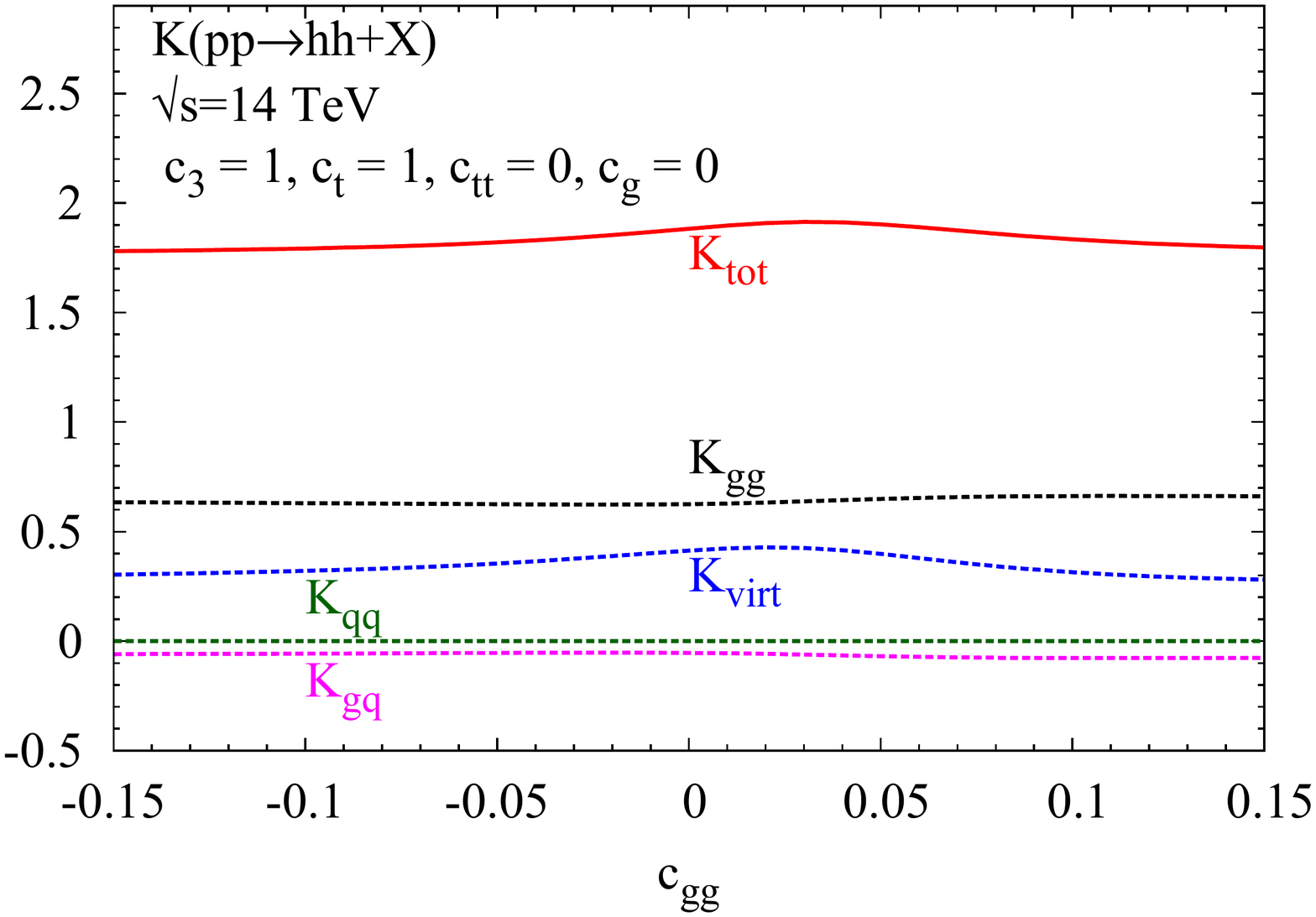}
\caption{$K$-factor as a function of $c_3$ (left) and $c_{gg}$ (right). The individual $K$-factors of the different contributions are defined as $K_i=\sigma_i/\sigma_{LO}$ with $i=virt,gg,gq, q\bar{q}$. \label{fig-2}}
\end{figure}
The results for $K=\sigma_{NLO}/\sigma_{LO}$ for a variation of $c_3$ and $c_{gg}$ are given in fig.~\ref{fig-2}.
From the figures it can be interferred that the SM $K$-factor approximates well the $K$-factor when dim 6 operators are included. The maximal difference to the SM $K$-factor is around 5\%. Note, however, that on the total cross section effective operators can have a large effect. 
The results of this subsection are publicly available in the code {\tt HPAIR} \cite{hpair}.  
\subsection{The MSSM}\label{MSSM}
This discussion summerizes the computation of the top and bottom squark contributions to the gluon fusion cross section to a pair of CP-even Higgs bosons in the MSSM of ref.~\cite{Agostini:2016vze}. 
\par
The QCD corrections for Higgs pair production via gluon fusion in the MSSM in the infinite top mass limit are known since long \cite{Dawson:1998py}. This computation includes, however, only the top and bottom contributions.
At LO, the contributions of top squarks and bottom squarks are given in refs.~\cite{Belyaev:1999mx}.
At two-loop order, one can make use of the triangle form factors computed in refs.~\cite{Anastasiou:2006hc} for single Higgs production.
The box form for the top/stop contributions can be computed via the low energy theorem (LET) for Higgs interactions \cite{Ellis:1975ap}. The LET connects the form factor $\mathcal{H}_{ij}^t$ (with $i$ and $j$ the indices of the Higgs field in the interaction basis) for the interaction of two Higgs bosons with two gluons at vanishing external momentum with the second derivative of the top/stop contributions to the gluon self-energy $\Pi^t$ with respect to the Higgs fields
\begin{equation}
\mathcal{H}_{ij}= \frac{2 \pi \, v^2}{\alpha_s T_F}\frac{\partial \Pi^t(0)}{\partial h_i \partial h_j}\, ,
\end{equation}
where $T_F=1/2$ is a color factor.
The sbottom contributions cannot be computed via the LET since the gluino contribution contains the bottom quark, whose mass is much smaller than the external momenta. The diagrams have hence to be computed explicitly via an asymptotic expansion in large sparticle masses. The limit of small external momenta used both for the top and bottom squark contributions is expected to work well for the pair production of a pair of SM-like Higgs bosons for diagrams with quartic squark couplings or gluons and squarks in realistic MSSM scenarios. Only contributions containing gluinos, top quarks and squarks will have, similar to the SM contribution, as discussed in section~\ref{SM}, thresholds at $m_{hh}= 2\,m_t$.
\par
The results of the computation have been implemented into a private version of {\tt HPAIR} \cite{hpair}. In order to exemplify the impact of the SUSY-QCD corrections, 
the parameter point
\begin{align}
&\tan \beta = 10,~~m_A=500~\text{GeV},~~\mu=-400~\text{GeV},
~~M_3=1500~\text{GeV},
\nonumber\\
&X_t=2\,M_S~,~~
m_{\tilde{t}_{\small{L}}}=m_{\tilde{t}_{\small{R}}}=m_{\tilde{b}_{\small{R}}}=M_S\,, \label{parapoint}
\end{align}
is chosen. In eq.~\eqref{parapoint}, $\tan\beta$ is the ratio of the vevs of the two Higgs doublets, $m_A$ the mass of the pseudoscalar Higgs boson, $m_{\tilde{t}_{\small{L}}}$, $m_{\tilde{t}_{\small{R}}}$ and $m_{\tilde{b}_{\small{R}}}$
are the soft-SUSY breaking masses of the squarks, $M_3$ the soft-SUSY breaking parameter of the gluino, $\mu$ the Higgs/higgsino mass term in the superpotential and $X_t=A_t+\mu \,\text{cot} \,\beta$ with $A_t$ denoting the soft SUSY breaking trilinear Higgs-top squark coupling. The squark mass scale $M_S$ is allowed to vary. 
\begin{figure}
\centering
\includegraphics[width=10cm]{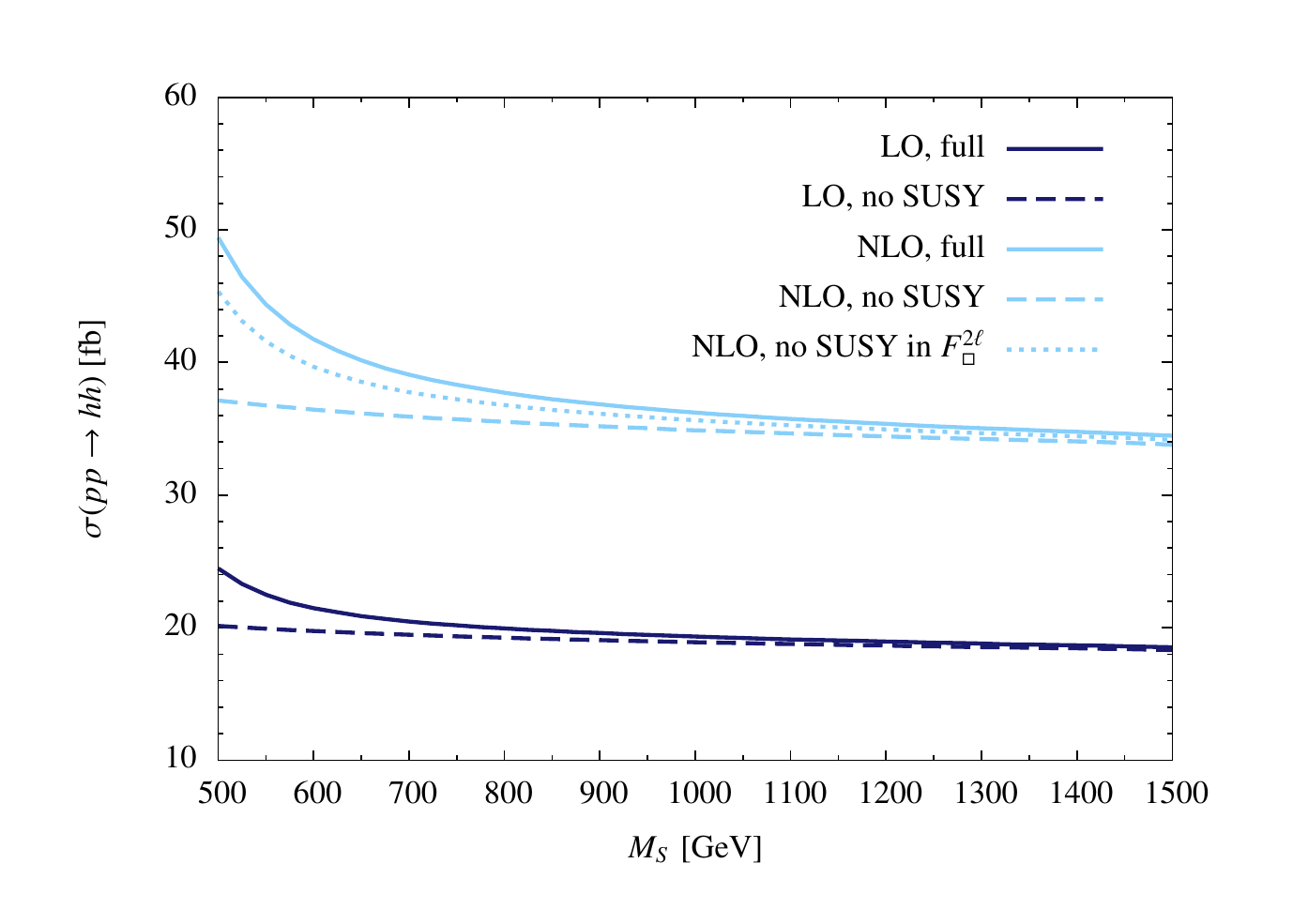}
\caption{Cross section for Higgs pair production via gluon fusion as a function of the squark mass scale. In dark-blue the LO cross section is shown, in light blue the NLO cross section. Dashed lines show the cross section without the stop/sbottom contributions, full lines include also SUSY contributions. The dotted dark-blue line does not include the two-loop SUSY contributions to the box form factor. \label{fig-MSSM}}
\end{figure}
The results for the Higgs pair production cross section of the SM-like Higgs boson can be found in fig.~\ref{fig-MSSM}. From the figure it can be inferred that the SUSY contributions for light squark mass scale increase the cross section by up to 30\%, whereas for large squark mass scale, the contributions of the stop/sbottoms become small. 
\section{Conclusion}\label{concl}
Higgs pair production is not only interesting as its measurement allows to determine the trilinear coupling but it can constrain new physics in many different ways.
In this contribution I have discussed whether it might be possible that new physics is for the first time observed in Higgs pair production in the context of Composite Higgs models.
Taking into account projected sensitivities on Higgs couplings' determination, projected sensitivity on direct searches for vector-like quarks and constraints from electroweak precision data and the measurement of $V_{tb}$,
it turns out that in a Composite Higgs model with both sizeable admixtures of bottom partners and top partners with the bottom or top quark, respectively, this can indeed be the case. The reason that the model can be distinguished from the SM can mainly be traced back to a large increase of the cross section due to a new $hht\bar{t}$ coupling.
\par
Higher order corrections to Higgs pair production via gluon fusion are numerically rather large. Hence it is not only important to know the cross section as precisely as possible in the SM, but also for SM extensions higher order corrections should be included. Here, I have discussed two examples: the SM extended with dim 6 operators and the MSSM. In the former case it turns out that the $K$-factor of the SM is a good approximation. For the MSSM instead, the SUSY contributions can be important for light squark masses.
\section{Acknowledgements}
I would like to thank the organizers of the Mini-Workshop on BSM Higgs for inviting me to such a nice and stimulating workshop in a very beautiful setting.
I am grateful to A.~Agostini, G.~Degrassi, P.~Giardino, M.~M\"uhlleitner, P.~Slavich, M.~Spira and J.~Streicher for discussion and collaboration on various projects on (BSM) Higgs pair production. In addition, I would like to thank M.~Spira for his comments on the manuscript.

\end{document}